\documentclass[11pt,a4paper]{article}
\usepackage[margin=2cm]{geometry}
\usepackage[utf8]{inputenc}
\usepackage[T1]{fontenc}
\usepackage{amsmath,amssymb,array,amsfonts}
\usepackage{booktabs}
\usepackage[sectionbib,round]{natbib}
\usepackage{graphicx}
\usepackage{fancyvrb}
\usepackage{alltt}
\RequirePackage[hyphens]{url}
\RequirePackage[pagebackref]{hyperref}

\RequirePackage{color}
\definecolor{link}{rgb}{0.45,0.51,0.67}
\hypersetup{
	colorlinks,%
	citecolor=link,%
	filecolor=link,%
	linkcolor=link,%
	urlcolor=link
}

\DeclareUnicodeCharacter{B1}{$\pm$}

\providecommand{\tightlist}{%
  \setlength{\itemsep}{0pt}\setlength{\parskip}{0pt}}

\newcommand{\CRANpkg}[1]{\href{https://CRAN.R-project.org/package=#1}{\pkg{#1}}}%
\newcommand{\dfn}[1]{{\normalfont\textsl{#1}}}
\newcommand{\strong}[1]{\texorpdfstring%
{{\normalfont\fontseries{b}\selectfont #1}}%
{#1}}
\let\pkg=\strong

\newcommand{\address}[1]{\addvspace{\baselineskip}\noindent\emph{#1}}

\DefineVerbatimEnvironment{Sinput}{Verbatim}{}
\DefineVerbatimEnvironment{Soutput}{Verbatim}{}
\DefineVerbatimEnvironment{Scode}{Verbatim}{}
\DefineVerbatimEnvironment{Sin}{Verbatim}{}
\DefineVerbatimEnvironment{Sout}{Verbatim}{}
\newenvironment{Schunk}{}{}

\makeatletter
\DeclareRobustCommand\code{\bgroup\@noligs\@codex}
\def\@codex#1{\texorpdfstring%
	{{\normalfont\ttfamily\hyphenchar\font=-1 #1}}%
	{#1}\egroup}
\providecommand{\operatorname}[1]{%
  \mathop{\operator@font#1}\nolimits}
\renewcommand{\P}{%
  \mathop{\operator@font I\hspace{-1.5pt}P\hspace{.13pt}}}
\newcommand{\E}{%
  \mathop{\operator@font I\hspace{-1.5pt}E\hspace{.13pt}}}

\makeatother

\begin{document}

\title{Measurement Errors in R}
\author{by Iñaki Ucar, Edzer Pebesma, Arturo Azcorra}

\maketitle

\abstract{%
This paper presents an R package to handle and represent measurements
with errors in a very simple way. We briefly introduce the main concepts
of metrology and propagation of uncertainty, and discuss related R
packages. Building upon this, we introduce the \CRANpkg{errors} package,
which provides a class for associating uncertainty metadata, automated
propagation and reporting. Working with \CRANpkg{errors} enables
transparent, lightweight, less error-prone handling and convenient
representation of measurements with errors. Finally, we discuss the
advantages, limitations and future work of computing with errors.
}

\hypertarget{introduction}{%
\section{Introduction}\label{introduction}}

The International Vocabulary of Metrology (VIM) defines a \dfn{quantity}
as ``a property of a phenomenon, body, or substance, where the property
has a magnitude that can be expressed as a number and a reference'',
where most typically the number is a \dfn{quantity value}, attributed to
a \dfn{measurand} and experimentally obtained via some measurement
procedure, and the reference is a \dfn{measurement unit}
\citep{VIM:2012}.

Additionally, any quantity value must accommodate some indication about
the quality of the measurement, a quantifiable attribute known as
\dfn{uncertainty}. The Guide to the Expression of Uncertainty in
Measurement (GUM) defines \dfn{uncertainty} as ``a parameter, associated
with the result of a measurement, that characterises the dispersion of
the values that could reasonably be attributed to the measurand''
\citep{GUM:2008}. Uncertainty can be mainly classified into
\dfn{standard uncertainty}, which is the result of a direct measurement
(e.g., electrical voltage measured with a voltmeter, or current measured
with a amperimeter), and \dfn{combined standard uncertainty}, which is
the result of an indirect measurement (i.e., the standard uncertainty
when the result is derived from a number of other quantities by the
means of some mathematical relationship; e.g., electrical power as a
product of voltage and current). Therefore, provided a set of quantities
with known uncertainties, the process of obtaining the uncertainty of a
derived measurement is called \dfn{propagation of uncertainty}.

Traditionally, computational systems have treated these three components
(quantity values, measurement units and uncertainty) separately. Data
consisted of bare numbers, and mathematical operations applied to them
solely. Units were just metadata, and uncertainty propagation was an
unpleasant task requiring additional effort and complex operations.
Nowadays though, many software libraries have formalised
\dfn{quantity calculus} as method of including units within the scope of
mathematical operations, thus preserving dimensional correctness and
protecting us from computing nonsensical combinations of quantities.
However, these libraries rarely integrate uncertainty handling and
propagation \citep{Flatter:2018}.

Within the R environment, the \CRANpkg{units} package
\citep{CRAN:units, Pebesma:2016:units} defines a class for associating
unit metadata to numeric vectors, which enables transparent quantity
derivation, simplification and conversion. This approach is a very
comfortable way of managing units with the added advantage of
eliminating an entire class of potential programming mistakes.
Unfortunately, neither \CRANpkg{units} nor any other package address the
integration of uncertainties into quantity calculus.

This article presents \CRANpkg{errors}, a package that defines a
framework for associating uncertainty metadata to R vectors, matrices
and arrays, thus providing transparent, lightweight and automated
propagation of uncertainty. This implementation also enables ongoing
developments for integrating units and uncertainty handling into a
complete solution.

\hypertarget{propagation-of-uncertainty}{%
\section{Propagation of uncertainty}\label{propagation-of-uncertainty}}

There are two main methods for propagation of uncertainty: the
\dfn{Taylor series method} (TSM) and the \dfn{Monte Carlo method} (MCM).
The TSM, also called the \dfn{delta method}, is based on a Taylor
expansion of the mathematical expression that produces the output
variables. As for the MCM, it is able to deal with generalised input
distributions and propagates the uncertainty by Monte Carlo simulation.

\hypertarget{taylor-series-method}{%
\subsection{Taylor series method}\label{taylor-series-method}}

The TSM is a flexible and simple method of propagation of uncertainty
that offers a good degree of approximation in most cases. In the
following, we will provide a brief description. A full derivation,
discussion and examples can be found in \citet{Arras:1998}.

Mathematically, an indirect measurement is obtained as a function of
\(n\) direct or indirect measurements, \(Y = f(X_1, ..., X_n)\), where
the distribution of \(X_n\) is unknown \emph{a priori}. Usually, the
sources of random variability are many, independent and probably unknown
as well. Thus, the central limit theorem establishes that an addition of
a sufficiently large number of random variables tends to a normal
distribution. As a result, the \strong{first assumption} states that
\(X_n\) are normally distributed.

The \strong{second assumption} presumes linearity, i.e., that \(f\) can
be approximated by a first-order Taylor series expansion around
\(\mu_{X_n}\) (see Figure \ref{propagation}). Then, given a set of \(n\)
input variables \(X\) and a set of \(m\) output variables \(Y\), the
first-order \dfn{uncertainty propagation law} establishes that

\begin{equation}\Sigma_Y = J_X \Sigma_X J_X^T\label{eq:prop-law}\end{equation}

\noindent where \(\Sigma\) is the covariance matrix and \(J\) is the
Jacobian operator.

\begin{Schunk}
\begin{figure}

{\centering \includegraphics[width=0.5\linewidth]{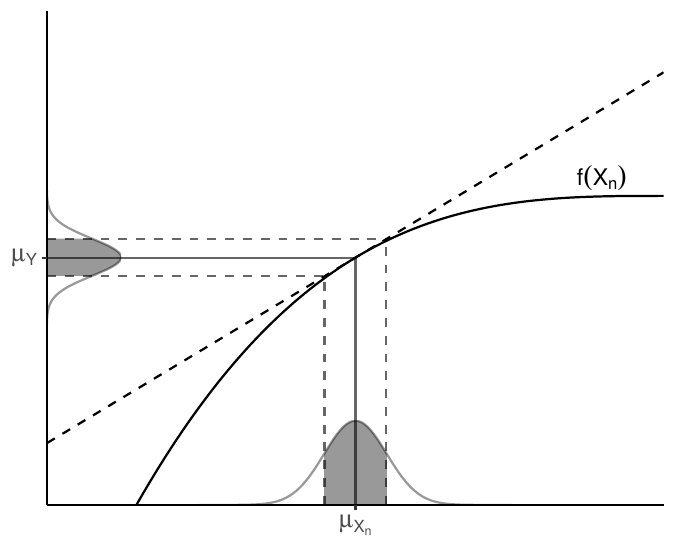} 

}

\caption[Illustration of linearity in an interval $\pm$ one standard deviation around the mean]{Illustration of linearity in an interval $\pm$ one standard deviation around the mean.\label{propagation}}\label{fig:fig1}
\end{figure}
\end{Schunk}

In practice, as recommended in the GUM \citep{GUM:2008}, this
first-order approximation is good even if \(f\) is non-linear, provided
that the non-linearity is negligible compared to the magnitude of the
uncertainty, i.e., \(\E[f(X)]\approx f(\E[X])\). Also, this weaker
condition is distribution-free: no assumptions are needed on the
probability density functions (PDF) of \(X_n\), although they must be
reasonably symmetric.

If we consider Equation \eqref{eq:prop-law} for pairwise computations,
i.e., \(Y = f(X_1, X_2)\), we can write the propagation of the
uncertainty \(\sigma_Y\) as follows:

\begin{equation}\label{eq:prop-var}
  \sigma_Y^2 = \left(\frac{\partial^2 f}{\partial X_1^2}\right)^2 \sigma_{X_1}^2 +
               \left(\frac{\partial^2 f}{\partial X_2^2}\right)^2 \sigma_{X_2}^2 +
               2 \frac{\partial f \partial f}{\partial X_1 \partial X_2} \sigma_{X_1 X_2}
\end{equation}

The cross-covariances for the output \(Y\) and any other variable \(Z\)
can be simplified as follows:

\begin{equation}\label{eq:prop-covar}
  \sigma_{Y Z} = \frac{\partial f}{\partial X_1} \sigma_{X_1 Z} +
                 \frac{\partial f}{\partial X_2} \sigma_{X_2 Z}
\end{equation}

\noindent where, notably, if \(Z=X_i\), one of the covariances above
results in \(\sigma_{X_i}^2\). Finally, and for the sake of
completeness, the correlation coefficient can be obtained as
\(r_{Y Z} = \sigma_{Y Z} / (\sigma_{Y}\sigma_{Z})\).

\hypertarget{monte-carlo-method}{%
\subsection{Monte Carlo method}\label{monte-carlo-method}}

The MCM is based on the same principles underlying the TSM. It is based
on the propagation of the PDFs of the input variables \(X_n\) by
performing random sampling and evaluating them under the model
considered. Thus, this method is not constrained by the TSM assumptions,
and explicitly determines a PDF for the output quantity \(Y\), which
makes it a more general approach that applies to a broader set of
problems. For further details on this method, as well as a comparison
with the TSM and some discussion on the applicability of both methods,
the reader may refer to the Supplement 1 of the GUM \citep{GUM:2008}.

\hypertarget{reporting-uncertainty}{%
\section{Reporting uncertainty}\label{reporting-uncertainty}}

The GUM \citep{GUM:2008} defines four ways of reporting standard
uncertainty and combined standard uncertainty. For instance, if the
reported quantity is assumed to be a mass \(m_S\) of nominal value 100
g:

\begin{quote}
\begin{enumerate}
\def\labelenumi{\arabic{enumi}.}
\tightlist
\item
  \(m_S = 100.02147\) g with (a combined standard uncertainty) \(u_c\) =
  0.35 mg.
\item
  \(m_S = 100.02147(35)\) g, where the number in parentheses is the
  numerical value of (the combined standard uncertainty) \(u_c\)
  referred to the corresponding last digits of the quoted result.
\item
  \(m_S = 100.02147(0.00035)\) g, where the number in parentheses is the
  numerical value of (the combined standard uncertainty) \(u_c\)
  expressed in the unit of the quoted result.
\item
  \(m_S = (100.02147 \pm 0.00035)\) g, where the number following the
  symbol \(\pm\) is the numerical value of (the combined standard
  uncertainty) \(u_c\) and not a confidence interval.
\end{enumerate}
\end{quote}

Schemes (2, 3) and (4) would be referred to as \dfn{parenthesis}
notation and \dfn{plus-minus} notation respectively throughout this
document. Although (4) is a very extended notation, the GUM explicitly
discourages its use to prevent confusion with confidence intervals.

\hypertarget{related-work}{%
\section{Related work}\label{related-work}}

Several R packages are devoted to or provide methods for propagation of
uncertainties. The \CRANpkg{car} \citep{CRAN:car, Fox:2011:car} and
\CRANpkg{msm} \citep{CRAN:msm, Jackson:2011:msm} packages provide the
functions \code{deltaMethod()} and \code{deltamethod()} respectively.
Both of them implement a first-order TSM with a similar syntax,
requiring a formula, a vector of values and a covariance matrix, thus
being able to deal with dependency across variables.

\begin{Schunk}
\begin{Sinput}
library(msm)

vals <- c(5, 1)
err <- c(0.01, 0.01)
deltamethod(~ x1 / x2, vals, diag(err**2))
\end{Sinput}
\begin{Soutput}
#> [1] 0.0509902
\end{Soutput}
\end{Schunk}

The \CRANpkg{metRology} \citep{CRAN:metRology} and \CRANpkg{propagate}
\citep{CRAN:propagate} packages stand out as very comprehensive sets of
tools specifically focused on this topic, including both TSM and MCM.
The \CRANpkg{metRology} package implements TSM using algebraic or
numeric differentiation, with support for correlation. It also provides
a function for assessing the statistical performance of GUM uncertainty
(TSM) using attained coverage probability. The \CRANpkg{propagate}
package implements TSM up to second order. It provides a unified
interface for both TSM and MCM through the \code{propagate()} function,
which requires an expression and a data frame or matrix as input. In the
following example, the uncertainty of \(x/y\) is computed, where
\(x=5.00(1)\) and \(y=1.00(1)\). The first result corresponds to the TSM
method and the second one, to the MCM.

\begin{Schunk}
\begin{Sinput}
library(propagate); set.seed(42)

x <- c(5, 0.01)
y <- c(1, 0.01)
propagate(expression(x/y), data=cbind(x, y), type="stat", do.sim=TRUE)
\end{Sinput}
\begin{Soutput}
#>    Mean.1    Mean.2      sd.1      sd.2      2.5%     97.5% 
#> 5.0000000 5.0005000 0.0509902 0.0509952 4.9005511 5.1004489 
#>       Mean         sd     Median        MAD       2.5%      97.5% 
#> 5.00042580 0.05101762 4.99991050 0.05103023 4.90184577 5.10187805
\end{Soutput}
\end{Schunk}

Unfortunately, as in the case of \CRANpkg{car} and \CRANpkg{msm}, these
packages are limited to work only with expressions, which does not solve
the issue of requiring a separate workflow to deal with uncertainties.

The \CRANpkg{spup} package \citep{CRAN:spup} focuses on uncertainty
analysis in spatial environmental modelling, where the spatial
cross-correlation between variables becomes important. The uncertainty
is described with probability distributions and propagated using MCM.

Finally, the \CRANpkg{distr} package
\citep{CRAN:distr, Ruckdeschel:2006:distr} takes this idea one step
further by providing an S4-based object-oriented implementation of
probability distributions, with which one can operate arithmetically or
apply mathematical functions. It implements all kinds of probability
distributions and has more methods for computing the distribution of
derived quantities. Also, \CRANpkg{distr} is the base of a whole family
of packages, such as \CRANpkg{distrEllipse}, \CRANpkg{distrEx},
\CRANpkg{distrMod}, \CRANpkg{distrRmetrics}, \CRANpkg{distrSim} and
\CRANpkg{distrTeach}.

All these packages provide excellent tools for uncertainty analysis and
propagation. However, none of them addresses the issue of an integrated
workflow, as \CRANpkg{units} does for unit metadata by assigning units
directly to R vectors, matrices and arrays. As a result, units can be
added to any existing R computation in a very straightforward way. On
the other hand, existing tools for uncertainty propagation require
building specific expressions or data structures, and then some more
work to extract the results out and to report them properly, with an
appropriate number of significant digits.

\hypertarget{automated-uncertainty-handling-in-r-the-package}{%
\section{\texorpdfstring{Automated uncertainty handling in R: the
\CRANpkg{errors}
package}{Automated uncertainty handling in R: the  package}}\label{automated-uncertainty-handling-in-r-the-package}}

The \CRANpkg{errors} \citep{CRAN:errors} package aims to provide easy
and lightweight handling of measurement with errors, including
uncertainty propagation using the first-order TSM presented in the
previous section and a formally sound representation.

\hypertarget{package-description-and-usage}{%
\subsection{Package description and
usage}\label{package-description-and-usage}}

Standard uncertainties, can be assigned to numeric vectors, matrices and
arrays, and then all the mathematical and arithmetic operations are
transparently applied to both the values and the associated
uncertainties:

\begin{Schunk}
\begin{Sinput}
library(errors)

x <- 1:5 + rnorm(5, sd = 0.01)
y <- 1:5 + rnorm(5, sd = 0.02)
errors(x) <- 0.01
errors(y) <- 0.02
x; y
\end{Sinput}
\begin{Soutput}
#> Errors: 0.01 0.01 0.01 0.01 0.01
#> [1] 0.990148 1.993813 2.995589 3.995865 5.011481
\end{Soutput}
\begin{Soutput}
#> Errors: 0.02 0.02 0.02 0.02 0.02
#> [1] 1.014570 1.996277 2.959307 3.992147 5.013118
\end{Soutput}
\begin{Sinput}
(z <- x / y)
\end{Sinput}
\begin{Soutput}
#> Errors: 0.021616206 0.011190137 0.007630253 0.005605339 0.004459269
#> [1] 0.9759291 0.9987660 1.0122601 1.0009312 0.9996735
\end{Soutput}
\end{Schunk}

The \code{errors()} method assigns or retrieves a vector of
uncertainties, which is stored as an attribute of the class
\code{errors}, along with a unique object identifier:

\begin{Schunk}
\begin{Sinput}
str(x)
\end{Sinput}
\begin{Soutput}
#>  'errors' num [1:5] 0.99(1) 1.99(1) 3.00(1) 4.00(1) 5.01(1)
#>  - attr(*, "id")= chr "2cf4231b-4bc1-412c-96f4-66f31d7e0399"
#>   ..- attr(*, ".Environment")=<environment: 0x5631fb052f08> 
#>  - attr(*, "errors")= num [1:5] 0.01 0.01 0.01 0.01 0.01
\end{Soutput}
\end{Schunk}

Correlations (and thus covariances) between pairs of variables can be
set and retrieved using the \code{correl()} and \code{covar()} methods.
These correlations are stored in an internal hash table indexed by the
unique object identifier assigned to each \code{errors} object. If an
object is removed, its associated correlations are cleaned up
automatically.

\begin{Schunk}
\begin{Sinput}
correl(x, x) # one, cannot be changed
\end{Sinput}
\begin{Soutput}
#> [1] 1 1 1 1 1
\end{Soutput}
\begin{Sinput}
correl(x, y) # NULL, not defined yet
\end{Sinput}
\begin{Soutput}
#> NULL
\end{Soutput}
\begin{Sinput}
correl(x, y) <- runif(length(x), -1, 1)
correl(x, y)
\end{Sinput}
\begin{Soutput}
#> [1]  0.39668415 -0.03191595  0.91062883 -0.37620892  0.25118328
\end{Soutput}
\begin{Sinput}
covar(x, y)
\end{Sinput}
\begin{Soutput}
#> [1]  7.933683e-05 -6.383190e-06  1.821258e-04 -7.524178e-05  5.023666e-05
\end{Soutput}
\end{Schunk}

Internally, \CRANpkg{errors} provides S3 methods for the generics
belonging to the groups \code{Math} and \code{Ops}, which propagate the
uncertainty and the covariance using Equations \eqref{eq:prop-var} and
\eqref{eq:prop-covar} respectively.

\begin{Schunk}
\begin{Sinput}
z # previous computation without correlations
\end{Sinput}
\begin{Soutput}
#> Errors: 0.021616206 0.011190137 0.007630253 0.005605339 0.004459269
#> [1] 0.9759291 0.9987660 1.0122601 1.0009312 0.9996735
\end{Soutput}
\begin{Sinput}
(z_correl <- x / y)
\end{Sinput}
\begin{Soutput}
#> Errors: 0.017799486 0.011332199 0.004014688 0.006393033 0.003986033
#> [1] 0.9759291 0.9987660 1.0122601 1.0009312 0.9996735
\end{Soutput}
\end{Schunk}

Other many S3 methods are also provided, such as generics belonging to
the \code{Summary} group, subsetting operators (\code{[}, \code{[<-},
\code{[[}, \code{[[<-}), concatenation (\code{c()}), differentiation
(\code{diff}), row and column binding (\code{rbind}, \code{cbind}),
coercion to list, data frame and matrix, and more. Such methods mutate
the \code{errors} object, and thus return a new one with no correlations
associated. There are also \emph{setters} defined as an alternative to
the assignment methods (\code{set\_*()} instead of \code{errors<-},
\code{correl<-} and \code{covar<-}), primarily intended for their use in
conjunction with the pipe operator (\code{\%>\%}) from the
\CRANpkg{magrittr} \citep{CRAN:magrittr} package.

Additionally, other useful summaries are provided, namely, the mean, the
weighted mean and the median. The uncertainty of any measurement of
central tendency cannot be smaller than the uncertainty of the
individual measurements. Therefore, the uncertainty assigned to the mean
is computed as the maximum between the standard deviation of the mean
and the mean of the individual uncertainties (weighted, in the case of
the weighted mean). As for the median, its uncertainty is computed as
\(\sqrt{\pi/2}\approx1.253\) times the standard deviation of the mean,
where this constant comes from the asymptotic variance formula
\citep{Hampel:2011:Robust}.

It is worth noting that both values and uncertainties are stored with
all the digits. However, when a single measurement or a column of
measurements in a data frame are printed, there are S3 methods for
\code{format()} and \code{print()} defined to properly format the output
and display a single significant digit for the uncertainty. This default
representation can be overriden using the \code{digits} option, and it
is globally controlled with the option \code{errors.digits}.

\begin{Schunk}
\begin{Sinput}
# the elementary charge
e <- set_errors(1.6021766208e-19, 0.0000000098e-19)
print(e, digits = 2)
\end{Sinput}
\begin{Soutput}
#> 1.6021766208(98)e-19
\end{Soutput}
\end{Schunk}

The \dfn{parenthesis} notation, in which \emph{the number in parentheses
is the uncertainty referred to the corresponding last digits of the
quantity value} (scheme 2 from the GUM, widely used in physics due to
its compactness), is used by default, but this can be overridden through
the appropriate option in order to use the \dfn{plus-minus} notation
instead.

\begin{Schunk}
\begin{Sinput}
options(errors.notation = "plus-minus")
print(e, digits = 2)
\end{Sinput}
\begin{Soutput}
#> (1.6021766208 ± 0.0000000098)e-19
\end{Soutput}
\begin{Sinput}
options(errors.notation = "parenthesis")
\end{Sinput}
\end{Schunk}

Finally, \CRANpkg{errors} also facilitates plotting of error bars. In
the following, we first assign a 2\% of uncertainty to all the numeric
variables in the \code{iris} data set and then we plot it using base
graphics and \CRANpkg{ggplot2}
\citep{CRAN:ggplot2, Wickham:2009:ggplot2}. The result is shown in
Figure \ref{plot}.

\begin{Schunk}
\begin{Sinput}
library(dplyr)

iris.e <- iris %>%
  mutate_at(vars(-Species), funs(set_errors(., .*0.02)))

head(iris.e, 5)
\end{Sinput}
\begin{Soutput}
#>   Sepal.Length Sepal.Width Petal.Length Petal.Width Species
#> 1       5.1(1)     3.50(7)      1.40(3)    0.200(4)  setosa
#> 2       4.9(1)     3.00(6)      1.40(3)    0.200(4)  setosa
#> 3      4.70(9)     3.20(6)      1.30(3)    0.200(4)  setosa
#> 4      4.60(9)     3.10(6)      1.50(3)    0.200(4)  setosa
#> 5       5.0(1)     3.60(7)      1.40(3)    0.200(4)  setosa
\end{Soutput}
\end{Schunk}

\begin{Schunk}
\begin{Sinput}
plot(iris.e[["Sepal.Length"]], iris.e[["Sepal.Width"]], col=iris.e[["Species"]])
legend(6.2, 4.4, unique(iris.e[["Species"]]), col=1:length(iris.e[["Species"]]), pch=1)
\end{Sinput}
\end{Schunk}

\begin{Schunk}
\begin{Sinput}
library(ggplot2)

ggplot(iris.e, aes(Sepal.Length, Sepal.Width, color=Species)) + 
  geom_point() + theme_bw() + theme(legend.position=c(0.6, 0.8)) +
  geom_errorbar(aes(ymin=errors_min(Sepal.Width), ymax=errors_max(Sepal.Width))) +
  geom_errorbarh(aes(xmin=errors_min(Sepal.Length), xmax=errors_max(Sepal.Length)))
\end{Sinput}
\end{Schunk}

\begin{Schunk}
\begin{figure}
\includegraphics[width=\linewidth]{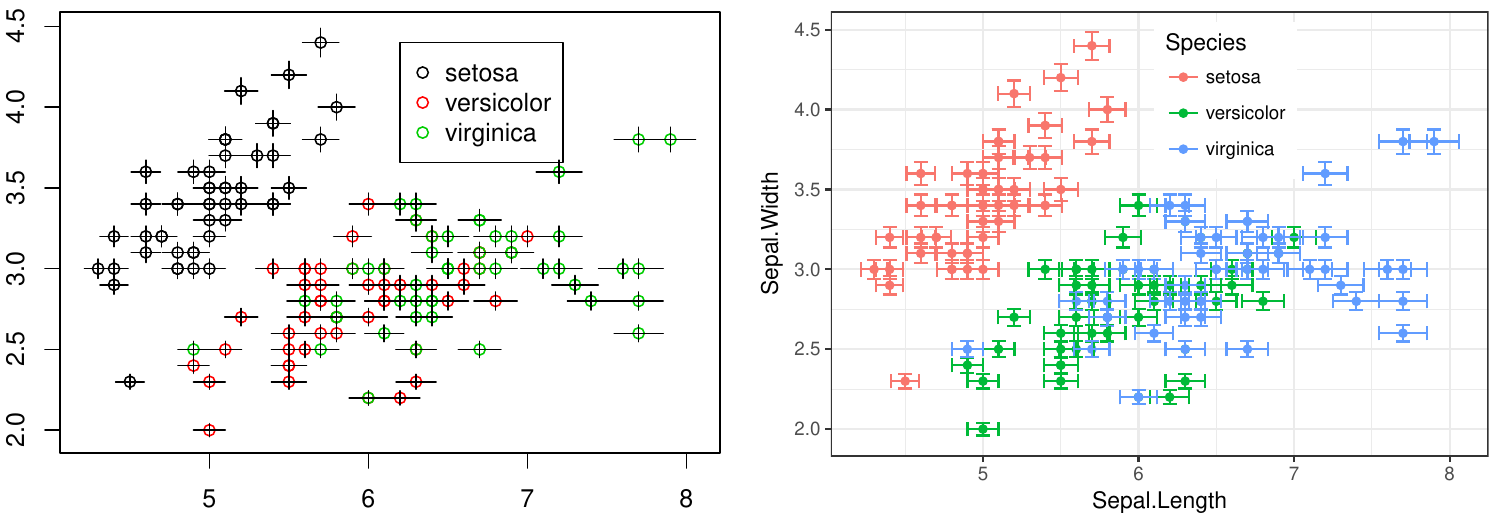} \caption{Base plot with error bars (left) and \CRANpkg{ggplot2}'s version (right).\label{plot}}\label{fig:fig2}
\end{figure}
\end{Schunk}

In base graphics, the error bars are automatically plotted when an
object of class \code{errors} is passed. Additionally, we provide the
convenience functions \code{errors\_min(x)} and \code{errors\_max(x)}
for obtaining the boundaries of the interval in \CRANpkg{ggplot2} and
other plotting packages, instead of writing \code{x - errors(x)} and
\code{x + errors(x)} respectively.

\hypertarget{example-simultaneous-resistance-and-reactance-measurement}{%
\subsection{Example: Simultaneous resistance and reactance
measurement}\label{example-simultaneous-resistance-and-reactance-measurement}}

From Annex H.2 of the GUM \citep{GUM:2008}:

\begin{quote}
The resistance \(R\) and the reactance \(X\) of a circuit element are
determined by measuring the amplitude \(V\) of a
sinusoidally-alternating potential difference across its terminals, the
amplitude \(I\) of the alternating current passing through it, and the
phase-shift angle \(\phi\) of the alternating potential difference
relative to the alternating current.
\end{quote}

The measurands (resistance \(R\), reactance \(X\) and impedance \(Z\))
are related to the input quantities (\(V\), \(I\), \(phi\)) by the Ohm's
law:

\begin{equation}
  R = \frac{V}{I}\cos\phi; \qquad X = \frac{V}{I}\sin\phi; \qquad Z = \frac{V}{I}
\end{equation}

Five simultaneous observations for each input variable are provided
(Table H.2 of the GUM), which are included in \CRANpkg{errors} as
dataset \code{GUM.H.2}. First, we need to obtain the mean input values
and set the correlations from the measurements. Then, we compute the
measurands and examine the correlations between them. The results agree
with those reported in the GUM:

\begin{Schunk}
\begin{Sinput}
V   <- mean(set_errors(GUM.H.2$V))
I   <- mean(set_errors(GUM.H.2$I))
phi <- mean(set_errors(GUM.H.2$phi))

correl(V, I)   <- with(GUM.H.2, cor(V, I))
correl(V, phi) <- with(GUM.H.2, cor(V, phi))
correl(I, phi) <- with(GUM.H.2, cor(I, phi))

print(R <- (V / I) * cos(phi), digits = 2, notation = "plus-minus")
\end{Sinput}
\begin{Soutput}
#> 127.732 ± 0.071
\end{Soutput}
\begin{Sinput}
print(X <- (V / I) * sin(phi), digits = 3, notation = "plus-minus")
\end{Sinput}
\begin{Soutput}
#> 219.847 ± 0.296
\end{Soutput}
\begin{Sinput}
print(Z <- (V / I), digits = 3, notation = "plus-minus")
\end{Sinput}
\begin{Soutput}
#> 254.260 ± 0.236
\end{Soutput}
\begin{Sinput}
correl(R, X); correl(R, Z); correl(X, Z)
\end{Sinput}
\begin{Soutput}
#> [1] -0.5884298
\end{Soutput}
\begin{Soutput}
#> [1] -0.4852592
\end{Soutput}
\begin{Soutput}
#> [1] 0.9925116
\end{Soutput}
\end{Schunk}

\hypertarget{example-calibration-of-a-thermometer}{%
\subsection{Example: Calibration of a
thermometer}\label{example-calibration-of-a-thermometer}}

From Annex H.3 of the GUM \citep{GUM:2008}:

\begin{quote}
A thermometer is calibrated by comparing \(n=11\) temperature readings
\(t_k\) of the thermometer {[}\ldots{}{]} with corresponding known
reference temperatures \(t_{R,k}\) in the temperature range 21
\(^\circ\)C to 27 \(^\circ\)C to obtain the corrections
\(b_k=t_{R,k}-t_k\) to the readings.
\end{quote}

Measured temperatures and corrections (Table H.6 of the GUM), which are
included in \CRANpkg{errors} as dataset \code{GUM.H.3}, are related by a
linear calibration curve:

\begin{equation}
  b(t) = y_1 + y_2 (t - t_0)
\end{equation}

In the following, we first fit the linear model for a reference
temperature \(t_0=20\) \(^\circ\)C. Then, we extract the coeficients,
and assign the uncertainty and correlation between them. Finally, we
compute the predicted correction \(b(30)\), which agrees with the value
reported in the GUM:

\begin{Schunk}
\begin{Sinput}
fit <- lm(bk ~ I(tk - 20), data = GUM.H.3)

y1 <- set_errors(coef(fit)[1], sqrt(vcov(fit)[1, 1]))
y2 <- set_errors(coef(fit)[2], sqrt(vcov(fit)[2, 2]))
covar(y1, y2) <- vcov(fit)[1, 2]

print(b.30 <- y1 + y2 * set_errors(30 - 20), digits = 2, notation = "plus-minus")
\end{Sinput}
\begin{Soutput}
#> -0.1494 ± 0.0041
\end{Soutput}
\end{Schunk}

\hypertarget{discussion}{%
\section{Discussion}\label{discussion}}

The \CRANpkg{errors} package provides the means for defining numeric
vectors, matrices and arrays with errors in R, as well as to operate
with them in a transparent way. Propagation of uncertainty implements
the commonly used first-order TSM formula from Equation
\eqref{eq:prop-law}. This method has been pre-computed and expanded for
each operation in the S3 groups \code{Math} and \code{Ops}, instead of
differentiating symbolic expressions on demand or using functions from
other packages for this task. The advantanges of this approach are
twofold. On the one hand, it is faster, as it does not involve
simulation nor symbolic computation, and very lightweight in terms of
package dependencies.

Another advantage of \CRANpkg{errors} is the built-in consistent and
formally sound representation of measurements with errors, rounding the
uncertainty to one significant digit by default and supporting two
widely used notations: \dfn{parenthesis} (e.g., \(5.00(1)\)) and
\dfn{plus-minus} (e.g., \(5.00 \pm 0.01\)). These notations are applied
for single numbers and data frames, as well as \code{tbl\_df} data
frames from the \CRANpkg{tibble} \citep{CRAN:tibble} package.

Full support is provided for both \code{data.frame} and \code{tbl\_df},
as well as matrices and arrays. However, some operations on those data
structures may drop uncertainties (i.e., object class and attributes).
More specifically, there are six common \emph{data wrangling}
operations: row subsetting, row ordering, column transformation, row
aggregation, column joining and (un)pivoting. Table \ref{tab:compat}
shows the correspondence between these operations and R base functions,
as well as the compatibility with \CRANpkg{errors}.

\begin{Schunk}
\begin{table}[t]

\caption{\label{tab:unnamed-chunk-14}Compatibility of \CRANpkg{errors} and R base data wrangling functions.\label{tab:compat}}
\centering
\begin{tabular}{lll}
\toprule
Operation & R base function(s) & Compatibility\\
\midrule
Row subsetting & \code{[}, \code{[[}, \code{subset} & Full\\
Row ordering & \code{order} + \code{[} & Full\\
Column transformation & \code{transform}, \code{within} & Full\\
Row aggregation & \code{tapply}, \code{by}, \code{aggregate} & with \code{simplify=FALSE}\\
Column joining & \code{merge} & Full\\
(Un)Pivoting & \code{reshape} & Full\\
\bottomrule
\end{tabular}
\end{table}

\end{Schunk}

Overall, \CRANpkg{errors} is fully compatible with data wrangling
operations embed in R base, and this is because those functions are
mainly based on the subsetting generics, for which \CRANpkg{errors}
provides the corresponding S3 methods. Nonetheless, special attention
must be paid to aggregations, which store partial results in lists that
are finally simplified. Such simplification is made with \code{unlist},
which drops all the input attributes, including custom classes. However,
all these aggregation functions provide the argument \code{simplify}
(sometimes \code{SIMPLIFY}), which if set to \code{FALSE}, prevents this
destructive simplification, and lists are returned. Such lists can be
simplified \emph{non-destructively} by calling \code{do.call(c, ...)}.

\begin{Schunk}
\begin{Sinput}
unlist <- function(x) if (is.list(x)) do.call(c, x) else x
iris.e.agg <- aggregate(. ~ Species, data = iris.e, mean, simplify=FALSE)
as.data.frame(lapply(iris.e.agg, unlist), col.names=colnames(iris.e.agg))
\end{Sinput}
\begin{Soutput}
#>      Species Sepal.Length Sepal.Width Petal.Length Petal.Width
#> 1     setosa       5.0(1)     3.43(7)      1.46(3)     0.25(1)
#> 2 versicolor       5.9(1)     2.77(6)      4.26(9)     1.33(3)
#> 3  virginica       6.6(1)     2.97(6)       5.6(1)     2.03(4)
\end{Soutput}
\end{Schunk}

\hypertarget{summary-and-future-work}{%
\section{Summary and future work}\label{summary-and-future-work}}

We have introduced \CRANpkg{errors}, a lightweight R package for
managing numeric data with associated standard uncertainties. The new
class \code{errors} provides numeric operations with automated
propagation of uncertainty through a first-order TSM, and a formally
sound representation of measurements with errors. Using this package
makes the process of computing indirect measurements easier and less
error-prone.

Future work includes importing and exporting data with uncertainties,
and providing the user with an interface for plugging uncertainty
propagation methods from other packages. Finally, \CRANpkg{errors}
enables ongoing developments for integrating \CRANpkg{units} and
uncertainty handling into a complete solution for quantity calculus.
Having a unified workflow for managing measurements with units and
errors would be an interesting addition to the R ecosystem with very few
precedents in other programming languages.

\bibliography{ucar-pebesma-azcorra}

\address{%
Iñaki Ucar\\
Universidad Carlos III de Madrid\\
Avda. de la Universidad, 30\\
28911 Leganés (Madrid), Spain\\
}
\href{mailto:inaki.ucar@uc3m.es}{\nolinkurl{inaki.ucar@uc3m.es}}

\address{%
Edzer Pebesma\\
Institute for Geoinformatics\\
Heisenbergstraße 2\\
48149 Münster, Germany\\
}
\href{mailto:edzer.pebesma@uni-muenster.de}{\nolinkurl{edzer.pebesma@uni-muenster.de}}

\address{%
Arturo Azcorra\\
Universidad Carlos III de Madrid\\
\emph{and}\\
IMDEA Networks Institute\\
Avda. de la Universidad, 30\\
28911 Leganés (Madrid), Spain\\
}
\href{mailto:azcorra@it.uc3m.es}{\nolinkurl{azcorra@it.uc3m.es}}

\end{document}